\documentclass[twocolumn,showpacs,preprintnumbers,amsmath,amssymb]{revtex4}

\usepackage{graphicx}
\usepackage{dcolumn}
\usepackage{bm}

\begin{document}

\title{\bf Microwave radiation induced magneto-oscillations in the longitudinal and
transverse resistance of a two dimensional electron gas }

\author{S.A. Studenikin\footnote{sergei.studenikin@nrc.ca},
M. Potemski$^\#$, P.T. Coleridge, A. Sachrajda, and Z.R.
Wasilewski} \affiliation{Institute for Microstructural Sciences,
National Research Council of Canada, Ottawa, Ontario, K1A OR6,
Canada\\
$^\#$Grenoble High Magnetic Field Laboratory, MPI/FKF and CNRS, BP
166, 38042 Grenoble, Cedex 9, France}
\date{7 October 2003 }
\begin{abstract}

We confirm the existance of magneto-resistance oscillations
in a microwave-irradiated two-dimensional
electron gas, first reported in a series of
papers by Zhudov et al.\cite{ZudovMani} and Mani et
al.\cite{ManiZudov}. In our experiments, on a sample with a more moderate
mobility, the microwave induced oscillations are observed not only
in the longitudinal - but also in the transverse-resistance (Hall
resistance). The phase of the oscillations is such that the
decrease (increase) in the longitudinal resistance is accompanied
by an increase (decrease) in the absolute value of the Hall
resistance. We believe that these new results
provide valuable new information to better understand the origin of this interesting phenomenon.

\end{abstract}

\pacs{ 73.50.Jt, 73.40.-c,78.67.-n, 78.20.Ls, 73.43.-f}

\keywords{microwaves,}

\maketitle
\newpage

\subsection{Introduction}

The new physics that results from subjecting a high mobility
two-dimensional electron gas (2DEG) to magnetic fields continues
to provide new and fascinating phenomena. Among them is the
recently reported observation of oscillations in the longitudinal
magneto-resistance ($\rho_{xx}$) of a 2DEG irradiated by
microwaves with frequencies in the range 5 - 100 GHz
\cite{ZudovMani}. These oscillations have been observed in samples
with very high mobility, at moderately low temperatures $T\sim1K$,
and at low magnetic fields  ($\hbar\omega_{c} \lesssim
\hbar\omega$). The periodicity of these oscillations is determined by the ratio of $\hbar\omega/\hbar\omega_{c}$, in
contrast to the well established case of Shubnikow-de Haas (SdH)
oscillations which they are determined by the ratio
$E_{F}/\hbar\omega_{c}$. Here, as usual: $\hbar\omega_{c}=\hbar
eB/m^{*}$ is the electron cyclotron energy, with $m^{*}$ the electron 
effective mass and $E_{F}$ denotes the Fermi energy. Under strong microwave irradiation and with samples that have a sufficiently high mobility, the minima in $\rho_{xx}$ become transformed into ``zero resistance states '' \cite{ManiZudov}.

It has been proposed \cite{Ryzhii,Durst,Shi} that these microwave
induced resistance oscillations (MIROs) can be understood in
terms of an additional, microwave-induced current which flows
with or against the direction of the applied electric field,
depending on whether $n-1/2 \lesssim \omega/\omega_{c} <
n$ or $n < \omega/\hbar\omega_{c} \lesssim n + 1/2$, with
$n=1,2,3,...$ integers. It has further been suggested that zero resistance states appear because of domain formation when the absolute conductivity becomes negative \cite{Andreev}. Alternatively, this effect may reflect even more elaborate physics \cite{ManiZudov}. A characteristic of the theoretical treatments is that no microwave induced changes to $\rho_{xx}$ are expected for $\omega/\omega_{c}$ exactly equal to 1,2,3,... \cite{Ryzhii,Durst,Shi}. Conventional cyclotron resonance absorption, which involves direct excitations and conservation of the
motion of the centre of the electron cyclotron orbit, does not result in any
additional current (i.e. does not change the conductivity). The additional
current appears as a result of indirect (off-diagonal) electron
excitations and involves other factors, such as impurities or phonons
\cite{Ryzhii,Durst,Shi}.  Therefore MIROs are not necessarily
triggered by a resonant absorption process, but more likely result from  non-resonant microwave-induced excitations of the 2DEG, with the modulation of $\rho_{xx}$ being a consequence of periodically conditions for the additional current to flow preferentially with
or against the applied electric field.

Diagrammatic calculations of the changes in the conductivity
tensor induced by the microwave-excited disorder-scattered
electrons reproduce the main experimental trends, in particular 
the observed period and the phase \cite{Durst}. Nevertheless, other
experimental features such as the exact shape of the oscillations as a function of magnetic field \cite{EP2DSNara} and the temperature dependence of the MIROs amplitudes are not as well understood. Another puzzle is that while the theory \cite{Durst} invokes an impurity-disorder mechanism to explain the MIROs, it is found experimentally that the oscillations become more clearly pronounced as the mobility is increased and the disorder is reduced.

In this paper we address another relevant issue, whether MIROs are a unique property of the longitudinal resistance (or alternatively $\sigma_{xx}$) or whether they also appear in the transverse resistance $\rho_{xy}$. So far, MIROs have been reported exclusively in measurements of  $\rho_{xx}$ or $\sigma_{xx}$ \cite{corbino}. It is sometimes argued that the absence of microwave induced features in $\rho_{xy}$ implies that 
MIROs have a classical origin rather than being a many-body effect. 
Theoretically, it has been predicted that both $\rho_{xx}$ and
$\rho_{xy}$ should be influenced by microwave irradiation
and that the resistance changes $\Delta\rho_{xx}$ and
$\Delta\rho_{xy}$, at least  under some approximations, should be comparable in magnitude \cite{Durst}. Results are presented here that show, for the first time, that microwave induced changes, of a similar magnitude, can be observed in measurements of both $\rho_{xx}$ and  $\rho_{xy}$ . It is shown that the observed oscillations in $\rho_{xy}$ cannot be attributed to more trivial effects, such as an admixture of $\rho_{xy}$
and $\rho_{xx}$ but rather must be associated with microwave induced oscillations in both the longitudinal conductivity $\sigma_{xx}$ and the transverse conductivity $\sigma_{xy}$. Possible consequences of our experimental findings are discussed.

\subsection{Experimental}

The measurements have been performed at He$^{4}$ bath temperatures
($T=1.4 - 4.2K$), on a GaAs/AlGaAs heterojunction (grown at NRC) that had
a 2DEG mobility of $\mu=4.0\times10^{6}cm^{2}/Vs$ and density of
$n=1.9\times10^{11}cm^{-2}$ after a brief illumination
using a red LED. The sample was cleaved into a rectangular shape to
form a 8x2 mm$^2$ Hall bar and contacted at the edges using small In dots. 
The length-to-width ratio between potential and Hall contacts was 1.3/2.

An Anritsu 69377B Signal Generator (operating at 0.01-50 GHz, with
typical output power of a few mW) was used as the source of
microwaves(MWs). They were delivered into the cryostat through a
semirigid 0.085 inch coaxial cable which had a typical attenuation factor 
of 6-10dB near 50GHz. The cable was terminated with a small antenna used to irradiate the sample which was mounted a few millimeters away. A small cylindrical cavity of MW absorbing material was constructed around the sample to suppress any cavity modes produced by metallic components of the cryostat.
The (perpendicular) magnetic fields produced by the superconducting magnet were carefully calibrated using  NMR and Hall effect probes and were checked
by measuring the EPR signal of DPPH powder mounted on a thermoresistor. 
Longitudinal and transverse resistances of the 2DEG were measured using 
standard  AC techniques, (at 15Hz and with measuring currents $\leq 1\mu A$). 
The data presented is the difference, $\Delta\rho$, between resistivities measured for two identical field sweeps with and without the application of MWs. 

\subsection{Results and discussion}

\begin{figure}[tbp]
\includegraphics[width=68mm,clip=false]{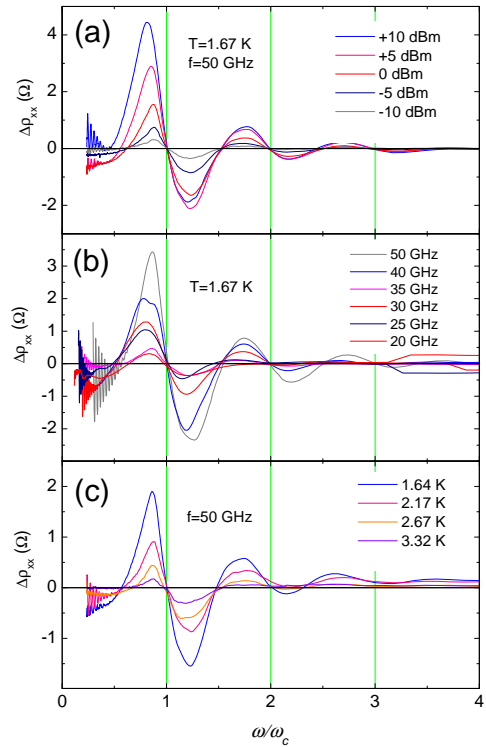}
\caption{Experimental traces of $\Delta\rho_{xx}$ induced by
microwave radiation plotted against normalized inverse magnetic
field $\omega/\omega_{c}$ ($\omega$ and $\omega_{c}$ is the
radiation and cyclotron cyclic frequency correspondingly) for
different powers (a), different frequencies (b), and for different
temperatures (c).} \label{fig1}
\end{figure}

Representative experimental traces of the longitudinal
magnetoresistance under microwave radiation are shown in Fig.1a
for different powers, Fig.1b for different frequencies, and Fig.1c
for different temperatures.  The data in Fig.1 is very similar that
previously reported for other MIROs \cite{ZudovMani,ManiZudov} (although the
mobility of the sample used here was too low to achieve ``zero resistance'' states), and confirms again that phenomena is universal, independent of the
source of the sample. Each $\Delta\rho_{xx}$ trace is plotted against 
the dimensionless parameter $\omega/\omega_{c}$, i.e., the ratio of microwave to cyclotron frequency, assuming the cyclotron mass, $m^{*}=0.067m_{e}$. In agreement with previous observations \cite{ZudovMani, ManiZudov}, $\Delta\rho_{xx}$ is found to be an oscillatory function of the
inverse magnetic field, with the fundamental period 
$\Delta(1/B)= (1/B)(\omega_c /\omega)$. It is apparent from Fig. 1 
that the oscillations are damped by increasing temperature
and by decreasing magnetic field, microwave power or
microwave frequency. The dependence of $\Delta\rho_{xx}$ on the
magnetic field can be roughly described as
\begin{equation}\label{eq.1}
\Delta\rho_{xx} = - A \exp(-D_{M}/\hbar\omega_{c}) \sin(2\pi\omega/ \omega_{c})
\end{equation}
with a phase corresponding to a positive coefficient A. The damping parameter $D_{M}$ is relatively insensitive to microwave power or frequency.  We
find that $D_{M}\approx$ 0.3meV at $T=1.64K$ and increases somewhat at higher temperatures. 
While Eq.(1) provides a good description of the higher order oscillations there are deviations around the fundamental cyclotron resonance
(n=1);  the exact positions of the maxima and minima in 
$\Delta\rho_{xx}$ become sensitive to the experimental parameters such as microwave power and temperature. This may be associated with the effect of the cyclotron resonance which can produce significant changes in the dielectric function of the 2DEG. The most robust and well defined feature of the MIROs are the positions of the zeros in $\Delta\rho_{xx}$ that occur when $\omega/\omega_{c}=1,2,3,...$. This is in agreement with theoretical predictions \cite{Ryzhii,Durst,Shi}.

The standard expression for the amplitude of the SdH oscillations is \cite{Coleridge}:
\begin{equation}\label{eq.2}
\Delta \rho_{xx} = 4 \rho_0 D_{th}(X_T) \exp(-\pi/\omega_c \tau_q) 
\end{equation}
where the thermal damping factor, $D_{th}$, is given by 
$X_T /\sinh(X_T)$ with $X_T = 2\pi^{2} kT / \hbar\omega_{c}$ and where $\tau_q$
is the quantum lifetime.

If the SdH oscillations in $\rho_{xx}$ (see for example figure 2) are analysed using this expression the damping is dominated by $D_{th}$ and no oscillations are visible in the low field region when the MIROs appear. However, measurements at lower temperatures (below 100mK) in another sample cut from the same wafer, show SdH oscillations to below .05 T which can be used to extract a value for $\tau_q$ of approximately 10ps. This corresponds to a damping coefficient
 $\pi \hbar/\tau_q$ of 0.27 meV, essentially identical to the coefficient $D_M$ obtained above for the MIROs.

At 0.1 tesla, $\tau_q$ = 10ps corresponds to a (Gaussian) Landau level broadening parameter of order 0.04 meV compared with the cyclotron spacing of 0.17 meV \cite{footnote2}. Therefore, despite the fact that thermal blurring of the distribution function suppresses the {\em observation} of the SdH oscillations at higher temperatures a well-defined Landau level structure still exists in the density of states. Furthermore, the damping of the MIROs appears to be the same as the disorder broadening of the Landau levels. This explains why MIROs can only be observed in very high mobility 2DEG samples where a well defined Landau level structure persists in the very low fields corresponding to $\omega_c/\omega \lesssim 1$.
 
For conventional SdH oscillations convolution of the Fermi function with the
density of states produces a strong thermal degradation of the oscillations
but this appears not to be the case for MIRO's. One explanation for this might be that the normal role of the Fermi function in producing a thermal smearing is masked by a stronger perturbation of the distribution function induced by the microwave radiation. It is only at higher temperatures (above about 1.5K) that the Fermi function can cause additional thermal smearing and give rise to a temperature dependent $D_M$.

\begin{figure}[tbp]
\includegraphics[width=68mm,clip=false]{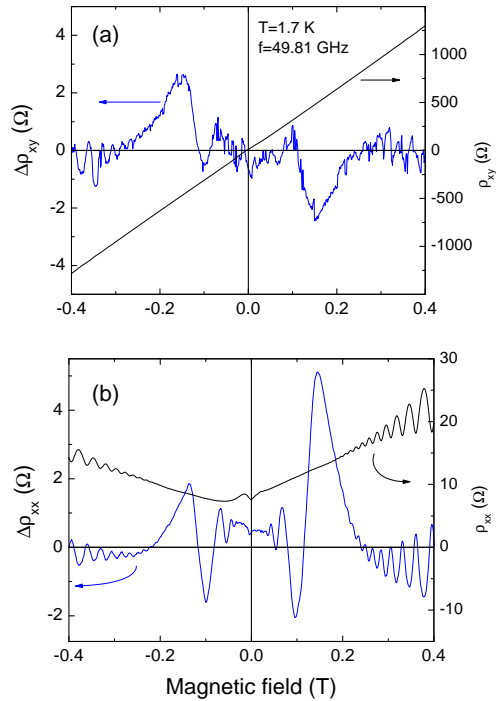}
\caption{Magnetic field dependencies of (a) the tranverse magnetoresistivity 
$\rho_{xy}$ and (b) the longitudinal magnetoresistivity $\rho_{xx}$ 
and their deviations from dark values induced
by microwave radiation f=49.8 GHz at T=1.7 K. } \label{fig2}
\end{figure}

In the following section we show that MIROs are not only evident in the
longitudinal resistance but can be also observed in the Hall
(transverse) resistance $\rho_{xy}$. This can be seen in Fig. 2,
which illustrates the result of simultaneous measurements of
$\rho_{xx}$ and $\rho_{xy}$ (right scale) and the MW induced deviations 
$\Delta\rho_{xx}$  and $\Delta\rho_{xy}$  (left axis). Despite the noise in the $\Delta\rho_{xy}$ trace it is clear that microwaves affect both resistances, and that the amplitudes of the two sets of oscillations are very similar. The noise is relatively large because the high mobility of the sample means $\Delta \rho_{xy}/\rho_{xy}$ is of order 100 times smaller than
$\Delta \rho_{xx}/\rho_{xx}$ when MIROs are observed so the inevitable errors associated with taking differences between large numbers become much more important for $\Delta \rho_{xy}$. 

Note that $\Delta \rho_{xy}$, like $\rho_{xy}$, changes sign in reversed magnetic fields. This means the observation of MIROs in $\rho_{xy}$ cannot be attributed to a trivial admixture of $\rho_{xx}$ into the $\rho_{xy}$ signal, due for example to misaligned contacts. The absolute sign of $\Delta \rho_{xy}$ depends on a sign convention but the fact that the sign of $\Delta \rho_{xy}/\rho_{xy}$ is the opposite to the sign of $\Delta \rho_{xx}$ is an unambiguous  experimental observation that can be tested against theory.

Simple quantitative arguments can also to made to confirm that $\Delta \rho_{xx}$ and $\Delta \rho_{xy}$  do not just reflect changes in $\sigma_{xx}$ exclusively. If the standard expressions are used for inverting the conductivities ({\it viz.} $\rho_{xx}=\sigma_{xx}/(\sigma_{xx}^{2}+\sigma_{xy}^{2})$ and
$\rho_{xy}=\sigma_{xy}/(\sigma_{xx}^{2}+\sigma_{xy}^{2})$)
then it is straightforward to show that a small change in $\Delta \sigma_{xx}$ 
will produce changes in $\rho_{xx}$ and $\rho_{xy}$ 
related by $\Delta \rho_{xy}= - \frac{2K}{1-K^{2}}\Delta \rho_{xx}$ where
K = $\rho_{xx}/\rho_{xy}$. While this gives the correct (observed) phase relationship between the two terms K is typically 0.02 so under the actual experimental conditions $\Delta \rho_{xy}$ is predicted 20 times smaller than $\Delta \rho_{xx}$, in sharp contrast to the
experimental results which show the ratio is actually about 2. 
An equivalent calculation shows, likewise, that the observed oscillations in $\Delta \rho_{xx}$ and $\Delta \rho_{xy}$ cannot result from changes in $\sigma_{xy}$ alone. We therefore conclude that both $\sigma_{xx}$ and $\sigma_{xy}$ are independently influenced by
microwaves. The results in Fig.2 showing $\Delta \rho_{xx}$ and $\Delta \rho_{xy}$ oscillating in anti-phase correspond, for example, to an increase (decrease) in effective electron concentration when the longitudinal conductivity ,$\sigma_{xx}$, increases (decreases). A theoretical determination of the phase relationship between $\Delta \rho_{xx}$ and $\Delta \rho_{xy}$ 
has not yet been addressed directly \cite{Durst}.

\subsection{Conclusions}
In conclusion, we have confirmed the observation of microwave
induced magneto-oscillations (MIROs) in the longitudinal
resistance of a 2DEG with a periodicity defined by the ratio of
microwave to cyclotron frequencies \cite{ZudovMani,ManiZudov}.

Our results indicate that a necessary condition for the
observation of these oscillations is a pronounced modulation of
the electronic density of states (Landau levels),  in
general agreement with predictions of existing theoretical models
\cite{Ryzhii,Durst,Shi}.

The temperature dependent damping of MIROs is very different from the 
thermal damping of the Shubnikov-de Haas oscillations. It is not clear whether this is related to a temperature dependence of the Landau level broadening or some other mechanism. Experiments to address this issue are in progress.

If MIROs are due to the scattering by impurities of  microwave-excited 
electrons but at the same time require a pronounced modulation in the density of states, a subtle criteria would exist for optimum sample quality. It would be interesting to check this if even higher mobility samples become available. However, phonon-assisted
processes may also play a role \cite{EP2DSNara} and effects resulting from the interplay between one-particle and collective excitations in the 2DEG \cite{Raman} should also not be neglected. More experiments, and in particular complementary microwave absorption measurements, are
needed to achieve a better understanding of this fascinating
phenomena.

The major new result reported here is that MIROs are not only
present in the longitudinal resistance but can also be observed in
the transverse component of the magneto-resistance. This is a non-trivial effect
and requires that $\sigma_{xx}$ and $\sigma_{xy}$ both oscillate independently. The
relative phase of the $\Delta \rho_{xx}$ and $\Delta \rho_{xy}$
oscillations experiment remains to be compared to theoretical predictions.

\subsection{Acknowledgements}

S.A.S and A.S. acknowledge support of The Canadian Institute for Advanced Research (CIAR).


\begin{thebibliography}{99}

\bibitem{ZudovMani} M.A. Zudov, R.R. Du, J.A. Simmons, and J.L. Reno, Phys. Rev. B \textbf{64}, 201311
(2001); R.G. Mani, J.H. Smet, K. von Klitzing, V. Narayanamurti,
W.B. Johnson, and V. Umansky, \textit{Proceedings of the $26^{th}$
International Conference on the Physics of Semiconductors}, (July
$29^{th}$-August $2^{nd}$, 2003, Edinbourg, U.K.), edited by A.C.
Long and J.H. Davies, (Institute of Physics), to be published;
cond-mat/0305507.

\bibitem{ManiZudov} R.G. Mani, J.H. Smet, K. von Klitzing, V. Narayanamurti,
W.B. Johnson, and V. Umansky, Nature (London) 420, 646 (2002);
M.A. Zudov, R.R. Du, J.A. Simmons, and J.L. Reno, Phys. Rev. B
Phys. Rev. Lett. \textbf{90}, 46807 (2003).

\bibitem{Ryzhii} V.I. Ryzhii, Fiz. Tverd. Tela \textbf{11},
2577(1969) [Sov. Phys. Solid State \textbf{11}, 2078 (1970)]; V.I.
Ryzhii\textit{et al.}, Fiz. Tekh. Poluprovodn. \textbf{20}, 2078
(1986) [Sov. Phys. Semicond. \textbf{20}, 1299 (1986)].

\bibitem{Durst}A.C. Durst, S. Sachdev, N. Read, and S.M. Girvin,
Phys. Rev. Lett. \textbf{91}, 86803 (2003).

\bibitem{Shi}J. Shi and X.C. Xie, Phys. Rev. Lett. \textbf{91}, 86805 (2003).

\bibitem{Andreev}A.V. Andreev, I.L. Aleiner, and A.J. Millis,
Phys. Rev. Lett. \textbf{91}, 56803, (2003).

\bibitem{EP2DSNara}For a recent review see special session at the EP2DS-Nara 2003 conference proceedings to be published.

\bibitem{corbino}  Corbino measurement?

\bibitem{footnote2} The Gaussian level broadening, determined according
to the self-consistent Born approximation, increases as B$^{1/2}$.

\bibitem{Raman}D. Richards, Phys. Rev. B \textbf{61}, 7517 (2000);
B. Jusserand, M. El Kurdi, and A. Cavanna, Phys. Rev. B \textbf{67}, 233307 (2003.

\bibitem{Coleridge} P. T. Coleridge, Phys. Rev. B \textbf{44}, 3793 (1991).

\end{thebibliography}
\end{document}